\documentclass[preprint]{aastex63}
\usepackage{subfigure}

\graphicspath{ {./figures/} }

\received{October 30, 2020}

\submitjournal{ApJ}

\shorttitle{Packing}
\shortauthors{Humphrey \& Quintana}

\graphicspath{{./}{figures/}}

\begin{document}

\title{Predicting missing planets in multiplanet system populations via analytical assessments of dynamical packing}

\correspondingauthor{Ana Luisa Ti\'{o} Humphrey}
\email{ana.humphrey@cfa.harvard.edu, elisa.quintana@nasa.gov}

\author[0000-0001-7072-0634]{Ana Luisa Ti\'{o} Humphrey}
\affiliation{Center for Astrophysics $\vert$ Harvard $\&$ Smithsonian \\
60 Garden Street \\
Cambridge MA 02138, USA}
\affiliation{T.C. Williams High School \\
3330 King Street \\
Alexandria VA 22302, USA}

\author{Elisa V. Quintana}
\affiliation{NASA Goddard Spaceflight Center \\
8800 Greenbelt Road \\
Greenbelt, MD 20771, USA}

\begin{abstract}

We present a new analytical method to identify potential missed planets in multi-planet systems found via transit surveys such as those conducted by Kepler and TESS. Our method depends on quantifying a system’s dynamical packing in terms of the dynamical spacing $\Delta$, the number of mutual Hill radii between adjacent planets (“planet pair”). The method determines if a planet pair within a multi-planet system is dynamically unpacked and thus capable of hosting an additional intermediate planet. If a planet pair is found to be unpacked, our method constrains the potential planet’s mass and location. We apply our method to the Kepler primary mission's population of 691 multi-candidate systems, first via direct calculations and then via Monte Carlo (MC) analysis. The analysis was repeated with three proposed values from the literature for minimum $\Delta$ required for planet pair orbital stability ($\Delta = 10$, $12.3$, and $21.7$). Direct calculations show that as many as $560$ planet pairs in 691 Kepler multi-candidate systems could contain additional planets ($\Delta = 12.3$). The MC analysis shows that $164$ of these pairs have a probability $\geq 0.90$ of being unpacked. Furthermore, according to calculated median mass efficiencies calculated from packed Kepler systems, $28.2\%$ of these potential planets could be Earths and Sub-Earths. If these planets exist, the masses and semimajor axes predicted here could facilitate detection by characterizing expected detection signals. Ultimately, understanding the dynamical packing of multi-planet systems could help contribute to our understanding of their architectures and formation.

\end{abstract}


\section{Introduction} \label{sec:intro}

More than $4100$ exoplanets have been confirmed to date by space missions such as Kepler and the Transiting Exoplanet Survey Satellite (TESS) as well as by ground-based surveys \citep{Borucki2003,Ricker2015}. The Kepler Space Telescope, launched in 2009, identified over $2300$ of these planets and $2200$ additional exoplanet candidates by detecting their transits across their host star. The more recent TESS mission has added over $60$ confirmed planets and $2100$ candidates to these totals. While detections via the transit method have been prolific, transit discoveries are limited by the difficulty of detecting smaller planets that do not block sufficient light from their host star and planets that do not orbit in our plane of view \citep{Borucki2003}. Kepler had difficulty detecting planets smaller than 1.2 Earth radii as they did not create a signal large enough be distinguishable from stellar photometric noise \citep{Howell2016}. This means that Kepler could not detect the Earth if it were to observe the Sun \citep{Howell2016}. However, investigations of exoplanet occurrence rates show the expected occurrence rate of planets increases as planet radius decreases (i.e. smaller planets should be the most common) \citep{Burke2015}. Thus, it is possible that many planets, especially those that are smaller or inclined, remain undetected in systems observed by Kepler. If such undetected low-mass planets exist, they could be invaluable to our understanding of system architectures and formation as well as the plausibility of life beyond earth.

\begin{figure}
    \centering
    \includegraphics[scale=.75]{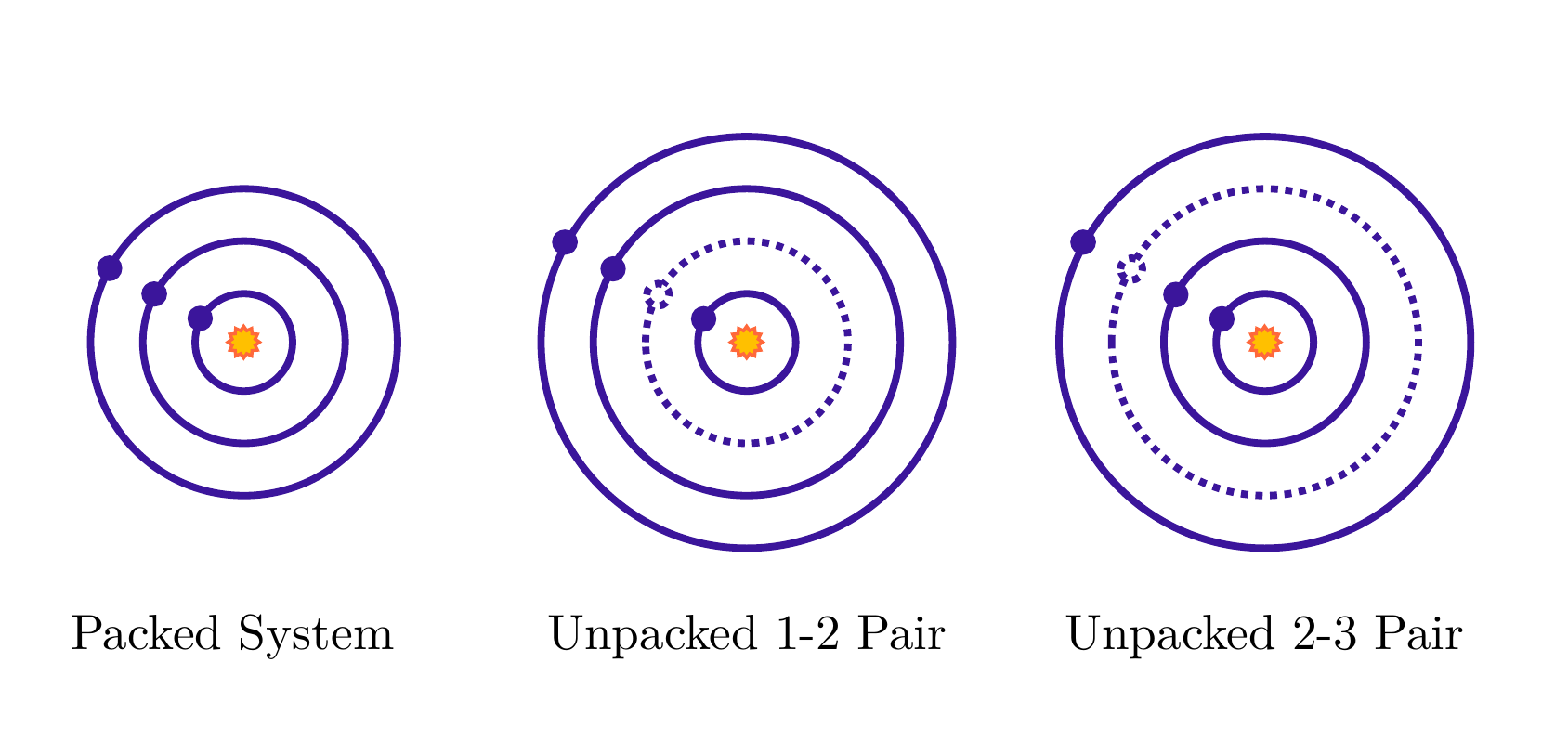}
    \caption{Examples of a packed and two three-planet unpacked systems. The middle system has an unpacked inner pair and a packed outer pair while the left system has a packed inner pair and an unpacked outer pair. According to PPS, unpacked planet pairs could contain an intermediate planet (indicated by dotted lines).}
    \label{fig:UnpackedPairDrawing}
\end{figure}

Analyses of the dynamical spacing of planets within the Kepler multi-planet extrasolar systems have suggested that some systems are capable of hosting additional exoplanets \citep{Gilbert2020,Kipping2018}. In conducting the analyses, researchers have evoked the Packed Planetary System (PPS) hypothesis, which postulates that the process of planetary formation is highly efficient, resulting in planetary systems that are dynamically packed and thus cannot contain additional planets \citep{Barnes2004}. Thus, identifying unpacked regions in multi-candidate systems should lead to the prediction of unidentified exoplanets (Figure \ref{fig:UnpackedPairDrawing}). 

The dynamical spacing between two planets in a system, denoted with $\Delta$, can be quantified in mutual Hill radii, a criterion that takes into account the distance between two adjacent planets in a planet pair, the planets’ masses, and the mass of the central star \citep{Gladman1993,Chambers1996}. The dynamical spacing between planets in a multiplanet system can be used to predict the system's stability \citep{Smith&Lissauer2009}. \cite{Chambers1996} found that adjacent planets in multiplanet systems whose dynamical spacing $\Delta$ is less than critical value $\Delta_{crit}=10$ are most likely unstable. In other words, planets that are closer than $10$ mutual Hill radii are likely to have close encounters, orbit crossing, and possible ejection from the system as a result of strong gravitational interaction between the planets.

\cite{Fang2012} investigated the stability of $115$ two-planet systems that had been identified by Kepler as of that year. By injecting massless test particles into the region between the pair of adjacent planets and performing numerical integrations with the modified systems for 107 years, \cite{Fang2012} determined that a planet pair with $\Delta < 10$ was unlikely to have a stability zone capable of containing a third planet in a stable orbit. These results confirmed findings by  \cite{Chambers1996} that $\Delta=10$ was the minimum spacing required for stability in multiplanet systems.

\cite{Fang2013} continued their PPS research by investigating the underlying architecture of Kepler systems. They generated systems independent of detection bias by exposing various system architectures to simulated Kepler observations and comparing the “detections” to those present in the Kepler data \citep{Fang2013}. Within their unbiased systems, they found that $\geq31\%$ of two-planet, $\geq35\%$ of three-planet, and $g\geq45\%$ of four-planet systems were dynamically packed, supporting the PPS hypothesis \citep{Fang2013}. They found a likely mean underlying $\Delta$ of $21.7$ mutual Hill radii. While the mean $\Delta$ was larger than the minimum $\Delta$ confirmed in 2012, their distribution for mean $\Delta$ was large with $\sigma  = 9.5$. \citep{Fang2012,Fang2013}. Furthermore, planets smaller than $1.5 R_{\oplus}$ were not included in their unbiased systems; as a result, tighter configurations may be common as less massive planets require a smaller region of stability to remain in stable orbits. 

\cite{PuandWu2015} used long-term numerical integration to examine the minimum spacing $\Delta$ required for generated seven-planet systems to maintain stability. They determined that systems containing planets with circular and coplanar orbits remained stable if their initial spacing was $\Delta=~10$, once again supporting the conclusions set forth by Chambers et al. \citep{Chambers1996, PuandWu2015}. Systems generated with eccentricities and mutual inclinations drawn from Poisson distributions with $\sigma \leq 0.02$ and $\sigma \leq 5^{\circ}$, respectively, required a minimum $\Delta$ of $12.3$ \citep{PuandWu2015}.
After the original Kepler mission ended in 2013, it took multiple years to process the Kepler data in its entirety. In 2016, DR24 was released, the first data release to contain all 48 months of Kepler data and the first data set uniformly vetted by an automatic robotic vetting \citep{Coughlin2016}. DR24 added $237$ planet candidates, removed $118$ false positives, and identified $48$ new multiplanet systems \citep{Coughlin2016}. DR24 was followed by DR25, which published results from an improved Robovetter that increased transit detection rates and reduced false positive rates, adding another 219 Kepler candidates \citep{Thompson2018}. When tested with artificially simulated transits, the DR25 catalog was $85.2\%$ complete and $97\%$ reliable \citep{Thompson2018}. In other words, while more complete than past data releases, there is a possibility of some missed transits in favor of maintaining a low false positive rate.

Since the publication of DR24 and DR25, an analysis of the stability and dynamical packing of all Kepler multicandidate systems has not been completed for the purpose of identifying additional potential planets. Weiss et al. \citep{Weiss2018} analyzed the dynamical spacing of planet pairs in $355$ multiplanet Kepler systems but did not extend the analysis to the possibility of additional intermediate planets within the planet pairs.

While the simulation techniques used by \cite{Fang2012}, and \cite{PuandWu2015} are too computationally expensive to be applied across the entire population of Kepler multiplanet systems, their results for minimum dynamical spacing provide a foundation for an analytical investigation of the Kepler multiplanet systems’ ability to host additional intermediate planets. The research presented here applies the $\Delta_{crit}$ suggested by Fang and Margot and Pu and Wu to examine the dynamical packing of Kepler systems with two to six candidates from the most updated Kepler catalog and identify planet pairs capable of hosting additional intermediate planets. This research could aid in the discovery of new planets by identifying systems capable of hosting as-of-yet unidentified planets. Furthermore, an analysis of the dynamical packing of systems using updated Kepler data could corroborate the PPS hypothesis, providing further insight into the formation and architecture of extrasolar systems.

We describe our methods for finding gaps (unpacked planet pairs) in multiplanet systems capable of hosting an additional planet in Section \ref{subsec:packing1} and our application of these methods to Kepler multiplanet systems in Section \ref{subsec:packing2}. We further characterize these gaps and predict possible mass and semimajor axis distributions for potential planets in section \ref{sec:predicting}. Results are discussed in Section \ref{sec:discussion}, and conclusions are summarized in Section \ref{sec:Conclusion}.

\section{Finding and Characterizing Unpacked Planet Pairs} \label{sec:Packing}
\subsection{Finding Regions of Stability and Unpacked Planet Pairs Analytically} \label{subsec:packing1}
To identify planet pairs capable of hosting unidentified intermediate planets, the investigation first determined whether the pair would remain stable if an additional intermediate planet were injected. This was determined by manipulating the analytical definition for $\Delta_{crit}$ set forth by \cite{Fang2012}. To be stable, a pair of planets 1 and 2 must keep a minimum dynamical spacing $\Delta_{crit}$:

\begin{equation}
    \Delta_{crit} \leq \frac{2(a_2-a_1)}{a_2+a_1}\left(\frac{M_1+M_2}{3M_*}\right)^{-\frac{1}{3}}
\end{equation}

where $a_1$ and $a_2$ are the planets’ semimajor axes (the average distance from their host star), $M_1$ and $M_2$ are their masses, and $M_*$ is the host star’s mass \citep{Chambers1996,Fang2012,Fang2013,PuandWu2015}. Thus, to host an intermediate planet $x$, $\Delta_{1x}$ and $\Delta_{x2}$ must satisfy:

\begin{equation}
    \Delta_{1x} \leq \frac{2(a_x-a_1)}{a_x+a_1}\left(\frac{M_1+M_x}{3M_*}\right)^{-\frac{1}{3}}
\end{equation}

\begin{equation}
    \Delta_{x2} \leq \frac{2(a_2-a_x)}{a_2+a_x}\left(\frac{M_x+M_2}{3M_*}\right)^{-\frac{1}{3}}
\end{equation}

Solving for $a_x$ gives $a_{xmin}$ and $a_{xmax}$, the minimum and maximum semimajor axes an intermediate planet $x$ could have while remaining in a stable orbit between planets $1$ and $2$:

\begin{equation} \label{axmin}
    a_{xmin}=2a_1 \left(\frac{2}{\Delta_{crit}}\left(\frac{M_1+M_x}{3M_*}\right)^{-\frac{1}{3}}-1\right)^{-1}+a_1
\end{equation}

\begin{equation} \label{axmax}
    a_{xmax}=-2a_2 \left(\frac{2}{\Delta_{crit}}\left(\frac{M_x+M_2}{3M_*}\right)^{-\frac{1}{3}}+1\right)^{-1}+a_2
\end{equation}

\begin{figure}
    \gridline{\fig{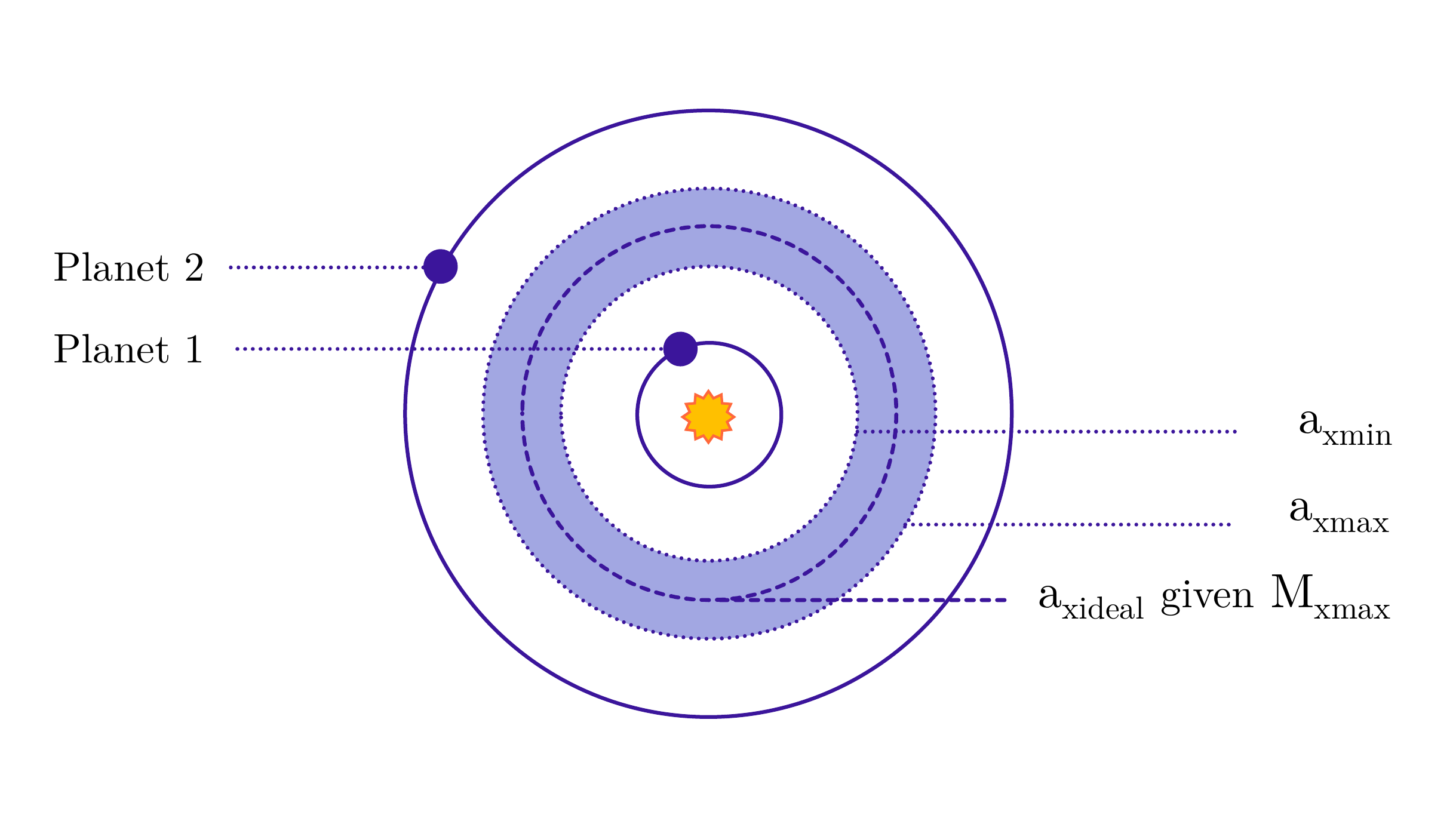}{.6\textwidth}{(a)}}
    \gridline{\fig{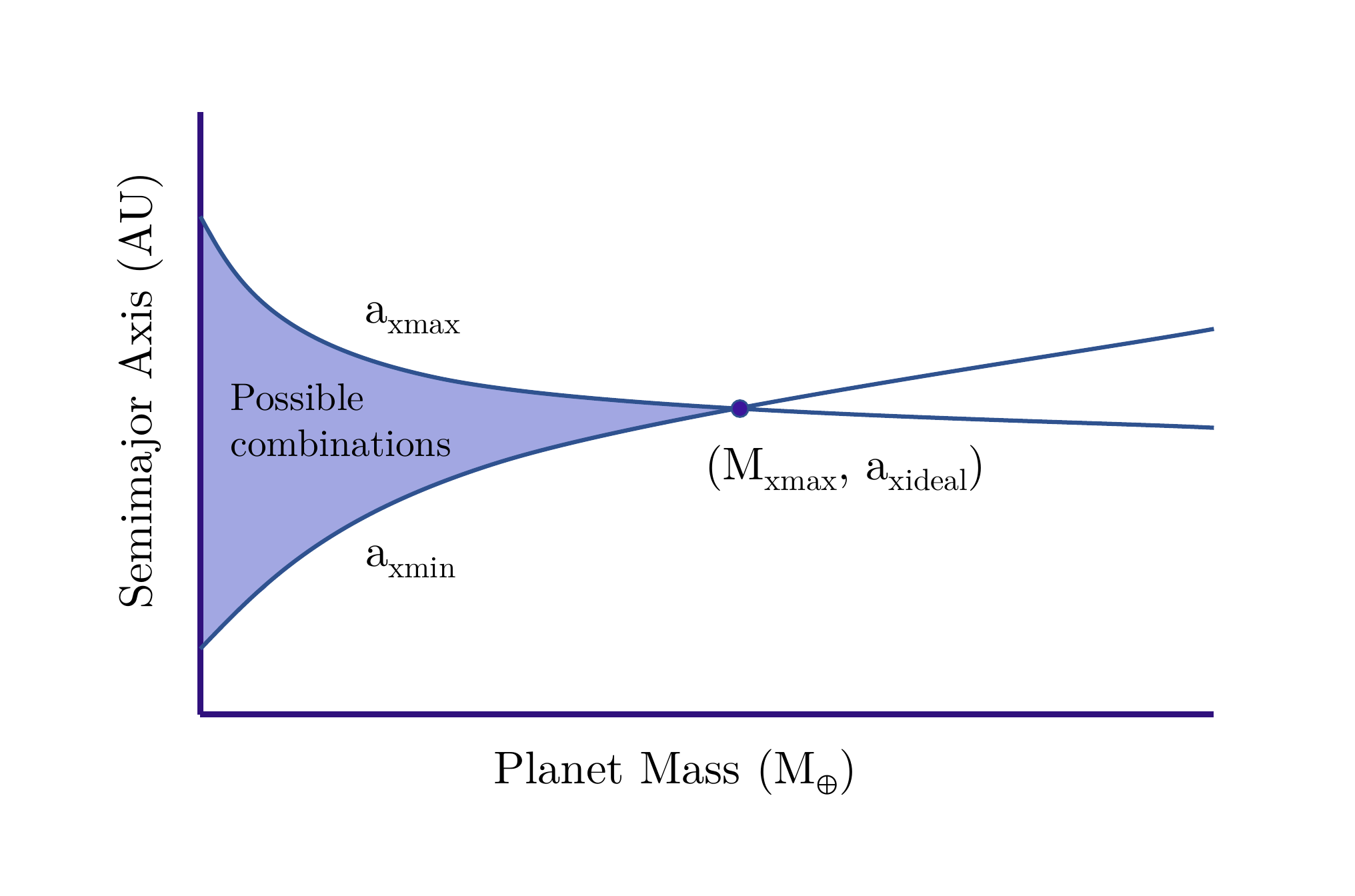}{.6\textwidth}{(b)}}
    \caption{(a) A planet pair capable of hosting an additional intermediate planet at a semimajor axis between $a_{xmin}$ and $a_{xmax}$. The shaded area between $a_{xmin}$ and $a_{xmax}$ represents the region of stability, which grows as the mass of the possible intermediate planet decreases. (b) A generic example of a solution to the system of equations \ref{axmin} and \ref{axmax}. Note that the x-axis represents mass and the y-axis represents semimajor axis for the potential intermediate planet. Equations \ref{axmin} ($a_{xmin}$) and \ref{axmax} ($a_{xmax}$) effectively show how close the intermediate planet's orbit can be to those of its inner and outer neighbors and remain stable as a function of the planet's possible mass. The shaded area between the curves indicates the possible mass and semimajor axis combinations the intermediate planet could have and remain in a stable orbit inside the planet pair. The solution to the system of equations represents the maximum mass, $M_{xmax}$, of an intermediate planet that the planet pair can contain and the ideal semimajor axis $a_{xideal}$ at which such a planet would orbit.}
    \label{fig:AnalyticalDemo}
\end{figure}

If $a_{xmin}\leq a_{xmax}$, then the pair is unpacked, i.e., sufficient dynamical spacing exists between planets $1$ and $2$ for the pair to host an intermediate planet x. The solution of the system of equations when $a_{xmin}=a_{xmax}$ provide the mass capacity $M_{xmax}$, the maximum mass that a hypothetical intermediate body could have without destabilizing the unpacked pair, and the ideal semimajor axis $a_{ideal}$, where the maximum mass could be held in a stable orbit (Figure \ref{fig:AnalyticalDemo}).
Three values were used for $\Delta_{crit}$: $10$, $12.3$, and $21.7$ to determine the sensitivity of the results to assumed minimum dynamical spacing of packed systems. $\Delta_{crit}=10$ was the minimum value suggested by \cite{Chambers1996} and \cite{Fang2012} for all systems and by \cite{PuandWu2015} for coplanar, circular systems. $12.3$ was the minimum value suggested by \cite{PuandWu2015} for systems with an eccentricity distribution of $\sigma \leq 0.02$ and mutual inclination distribution of $\sigma \leq 5^{\circ}$. $21.7$ was the mean value \cite{Fang2013} suggested for Kepler systems. 

\subsection{Constraining Potential Planets in the Kepler Multiplanet Systems}\label{subsec:packing2}
The full set of planet candidates in the Kepler DR25 catalog \citep{Thompson2018} was accessed via the NASA Exoplanet Archive\footnote{https://exoplanetarchive.ipac.caltech.edu/}. Of the 4496 planet candidates, 1715 are in multiple planet systems. The population of multiplanet systems we analyze thus include 460 2-planet systems, 157 3-planet systems, 50 4-planet systems, 20 5-planet systems, and 4 6-planet systems. 

\begin{figure}
    \gridline{\fig{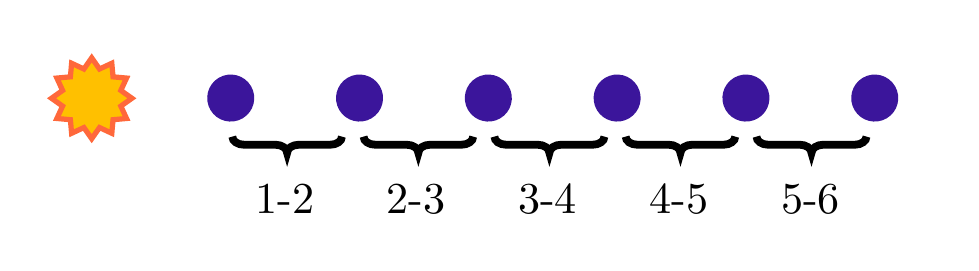}{.6\textwidth}{(a)}}
    \gridline{\fig{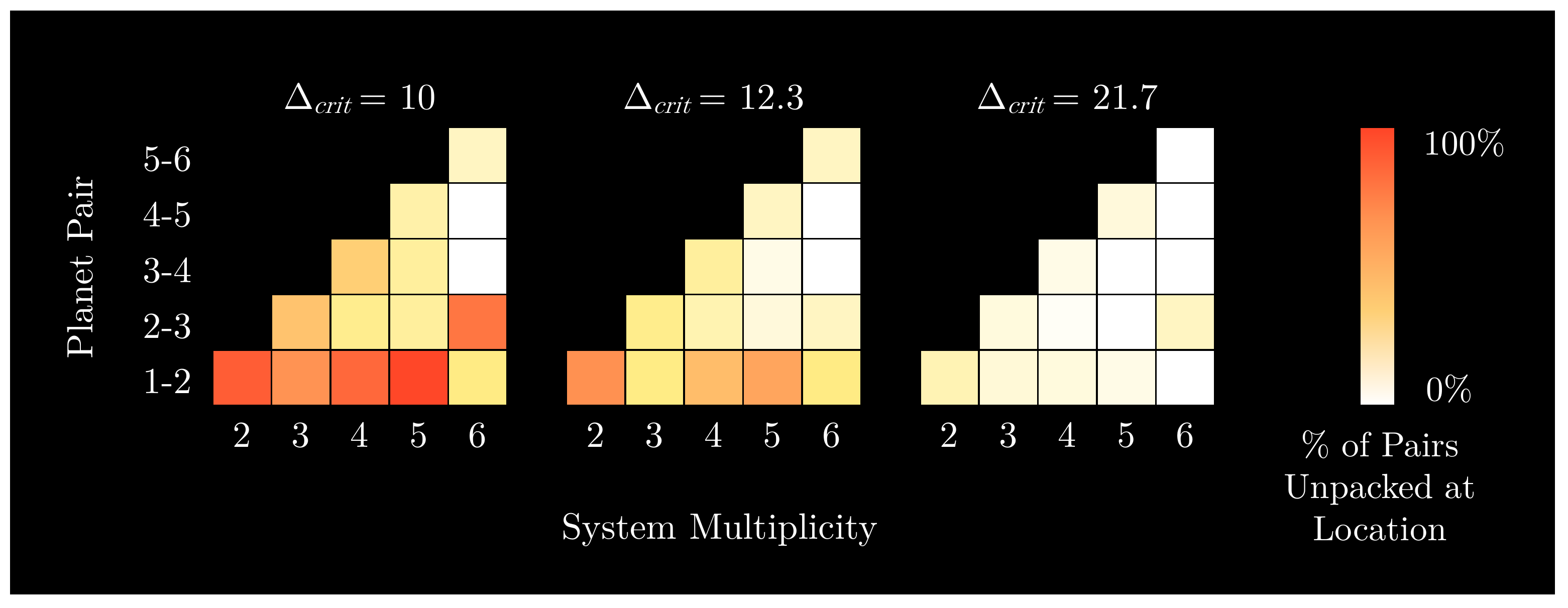}{1\textwidth}{(b)}}
    \caption{(a) The pair locations in a Kepler multi-candidate system. 1-2 pairs are closest to the host star. (b) The percentage of planet pairs in each pair location capable of hosting a potential planet in Kepler systems of different multiplicities. The color of each box corresponds to the percentage of planet pairs capable of hosting a potential planet. The colors scale from white to yellow to orange to red, with dark red indicating that a high percentage of pairs in that location can host a potential planet. The vertical axis denotes the system multiplicity while the horizontal axis denotes the pair’s location in the system.}
    \label{fig:UnpackedPairLocation}
\end{figure}

Transits can provide the radius but not the mass of a candidate. To obtain mass estimates for the planets, we used the empirical mass-radius software package Forecaster \citep{ChenandKipping2017}, which enabled predictions of the masses of the Kepler candidates based on their radii. Forecaster predicts a possible distribution of masses for each candidate radius based on a four-segment least-squares fit of $316$ well-constrained celestial objects.
All $1024$ pairs of adjacent planets in the Kepler multiplanet systems were examined analytically using each parameter’s central values in the KOI table. Solving the system of Equations \ref{axmin} and \ref{axmax} resulted in a binary indication of whether a pair was unpacked and produced a single $M_{xmax}-a_{ideal}$ solution per pair. The analysis of the Kepler multiplanet system population was repeated for each of the chosen values of $\Delta_{crit}$ ($10$, $12.3$, and $21.7$). The results indicate as many as $701$, $560$, and $206$ planets are possible according to $\Delta_{crit}= 10$, $12.3$, and $21.7$, respectively. The 1-2 planet pair, or the planet pair closest to the central star, was the most likely to be unpacked and thus capable of hosting an additional planet in a multiplanet system (Figure \ref{fig:UnpackedPairLocation}). However, with the exception of the six-planet systems, of which there were only four in the sample, it is possible to find an unpacked planet pair in any location in a system regardless of multiplicity.

\begin{figure}
    \centering
    \includegraphics[scale=.5]{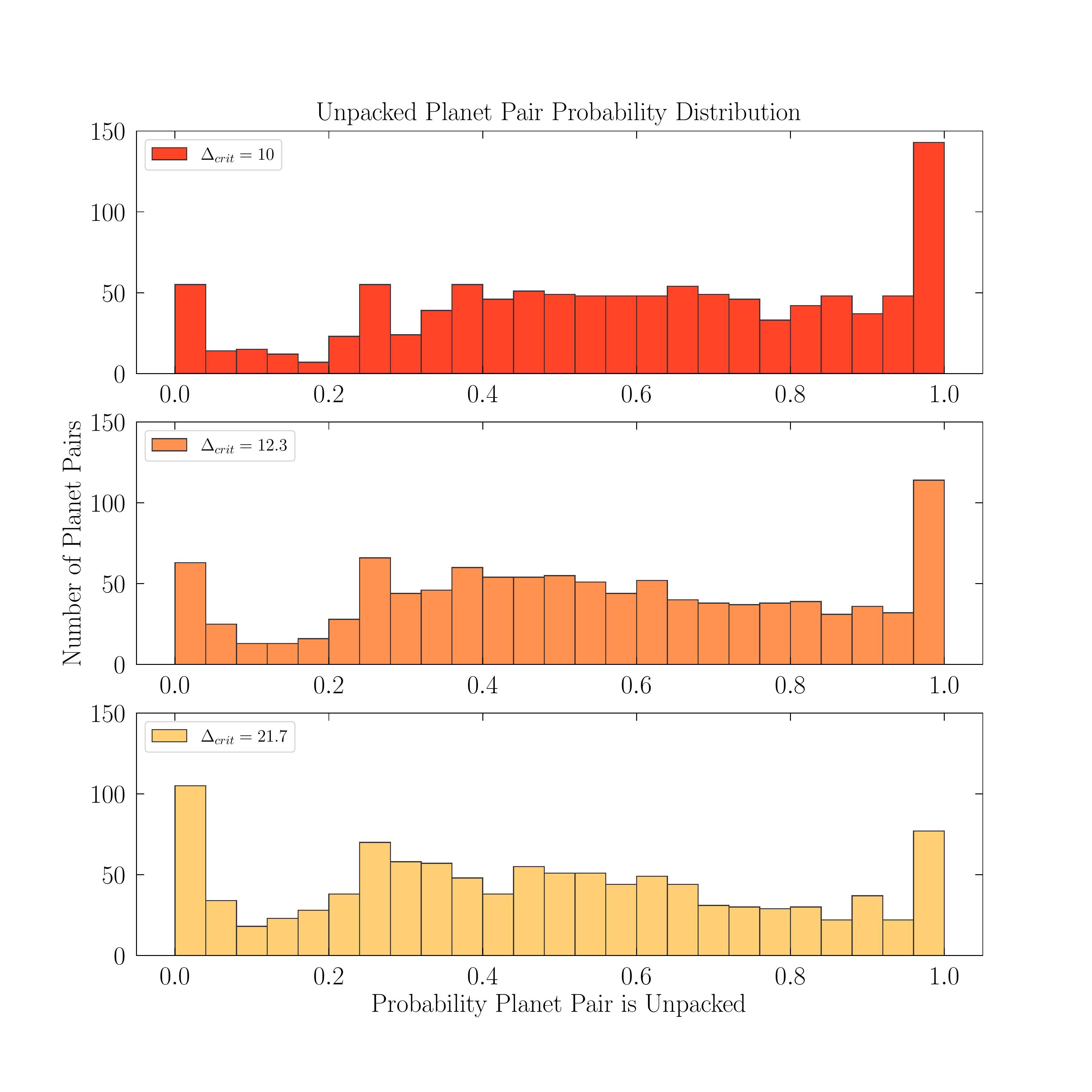}
    \caption{The MC distribution of planet pairs’ probabilities of being unpacked given the three chosen values for $\Delta_{crit}$.}
    \label{fig:UnpackedPlanetPairProb}
\end{figure}

To account for the uncertainties in the five core input parameters for each planet pair (planet masses, planet semimajor axis, and stellar mass), the analysis was repeated using Monte Carlo (MC) simulations. A posterior distribution was fit to each of the input parameters based on their measurement uncertainties. The system of Equations \ref{axmin} and \ref{axmax} was then solved $1000$ times for each planet pair with input parameters randomly sampled from each parameter’s distribution. This yielded probabilities that a planet pair was unpacked as well as output distributions for a pair’s $M_{xmax}$ and $a_{ideal}$. $209$, $164$, and $118$ planet pairs had a probability $\geq 0.90$ of being unpacked for $\Delta_{crit} = 10$, $12.3$, and $21.7$, respectively (Figure \ref{fig:UnpackedPlanetPairProb}). If unpacked planet pairs are assumed to contain additional undetected planets, these pairs represent the locations most likely to contain these additional planets.

\begin{figure}
    \centering
    \includegraphics[scale=0.5]{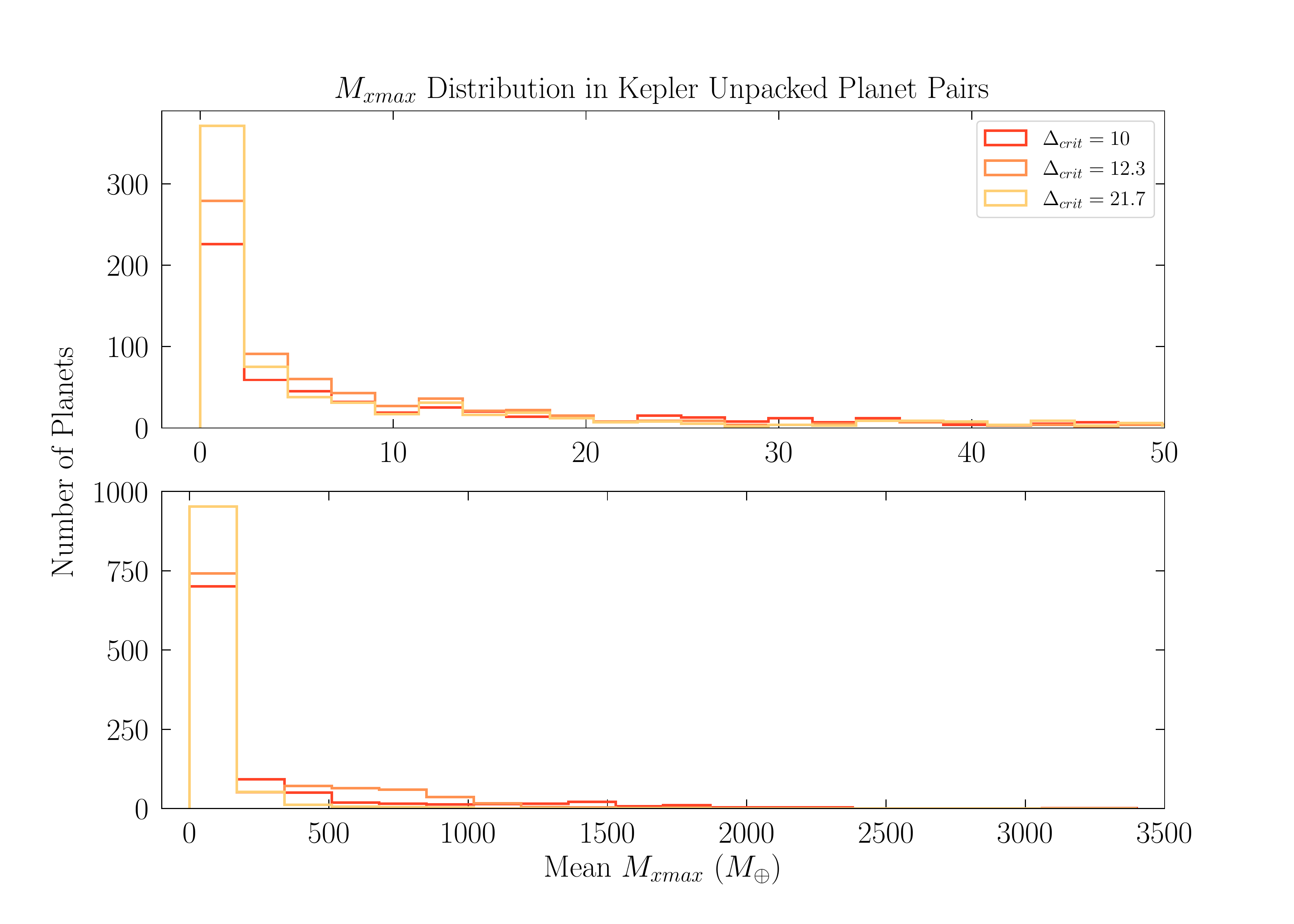}
    \caption{The distribution of mean $M_{xmax}$  for unpacked planet pairs given the three chosen values for $\Delta_{crit}$.}
    \label{fig:MaxMassDist}
\end{figure}

The posterior distribution of mass capacities for each unpacked planet pair indicate how massive of a planet the pair could accommodate. As the minimum assumed $\Delta_{crit}$ required for planets stability increases, the mass capacity of unpacked planet pairs decreases. As a result, 370, 487, and 523 planet pairs’ mean $M_{xmax}$ were $\leq 10 M_{\oplus}$ (Super-Earth or smaller) when $\Delta_{crit} = 10$, $12.3$, and $21.7$, respectively. While unpacked planet pairs were in general more likely to have a mass capacity of $10 M_{\oplus}$ or smaller, some planet pairs were fairly spaced apart and could host an additional planet with a mass up to or larger than Jupiter (Figure \ref{fig:MaxMassDist}).

\section{Predicting Planet Location and Mass}\label{sec:predicting}

To determine if the mass capacity ($M_{xmax}$) and ideal semimajor axis of a planet pair is indicative of an unidentified intermediate planet’s mass and semimajor axis ($a_{ideal}$), the analytical analysis was repeated for packed triplets, three consecutive planets in a system comprised of two packed planet pairs. A packed system with an unidentified intermediate planet was simulated by “removing” an intermediate planet in a packed triplet, creating an unpacked planet pair. $M_{xmax}$ and $a_{ideal}$ were calculated for the simulated unpacked planet pair and compared to the mass and semimajor axis of the removed planet. Mass was compared using the mass efficiency ($M_{eff}$), calculated by finding the ratio between the removed planet’s mass and the mass capacity of the planet pair formed by the remaining planets.

\begin{equation}
    M_{eff}=\frac{M_{intermediate}}{M_{xmax}}
\end{equation}

The semimajor axis was compared using the semimajor axis deviation ($a_{dev}$), the deviation between the intermediate planets’ semimajor axes ($a$) as published in the NASA Exoplanet Archive and their ideal semimajor axes ($a_{ideal}$):

\begin{equation}
    a_{dev}=\frac{a_{ideal}-a}{a}
\end{equation}

\begin{figure}
    \centering
    \includegraphics[scale=.5]{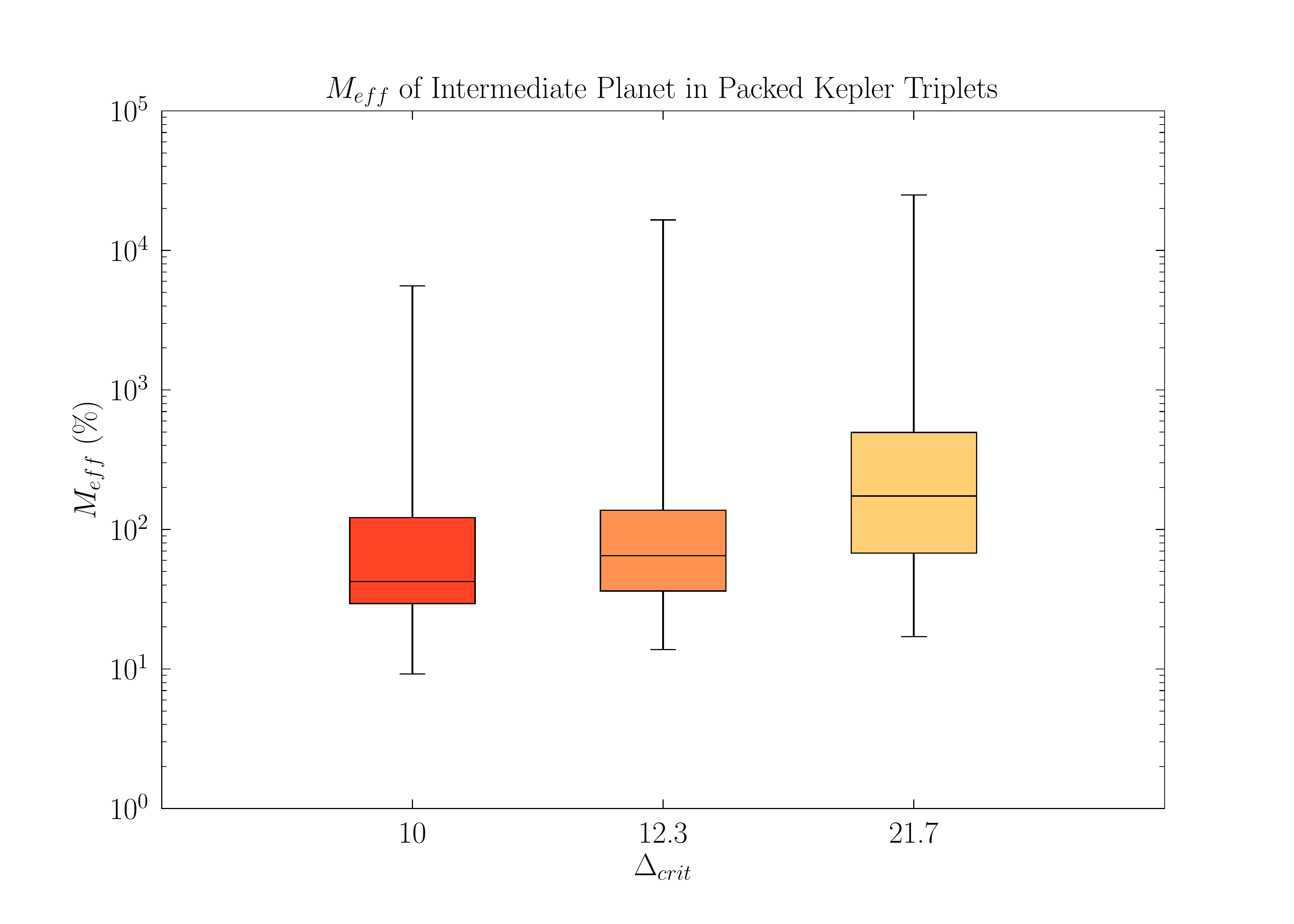}
    \caption{The range of mass efficiencies, given as a percentage, ($M_{eff}$) of intermediate planets in Kepler packed triplets for $\Delta_{crit} = 10$, $12.3$, and $21.7$. The central lines in each box indicates the median mass efficiency among the population of Kepler packed triplets given each value of $\Delta_{crit}$. The shaded boxes represent the mass efficiencies ranging from the 25th-50th and 50th-75th percentile given each $\Delta_{crit}$. The bars extending on either side of the boxes indicate the minimum and maximum $M_{eff}$. The median (50th-percentile) $M_{eff}$ were $42.28\%$, $64.93\%$, and $173.56\%$, respectively. Some $M_{eff}$ exceeded $100\%$, suggesting the systems are unstable or Forecaster masses are too large.}
    \label{fig:PackedTripletMassEfficiency}
\end{figure}

\begin{figure}
    \centering
    \includegraphics[scale=0.5]{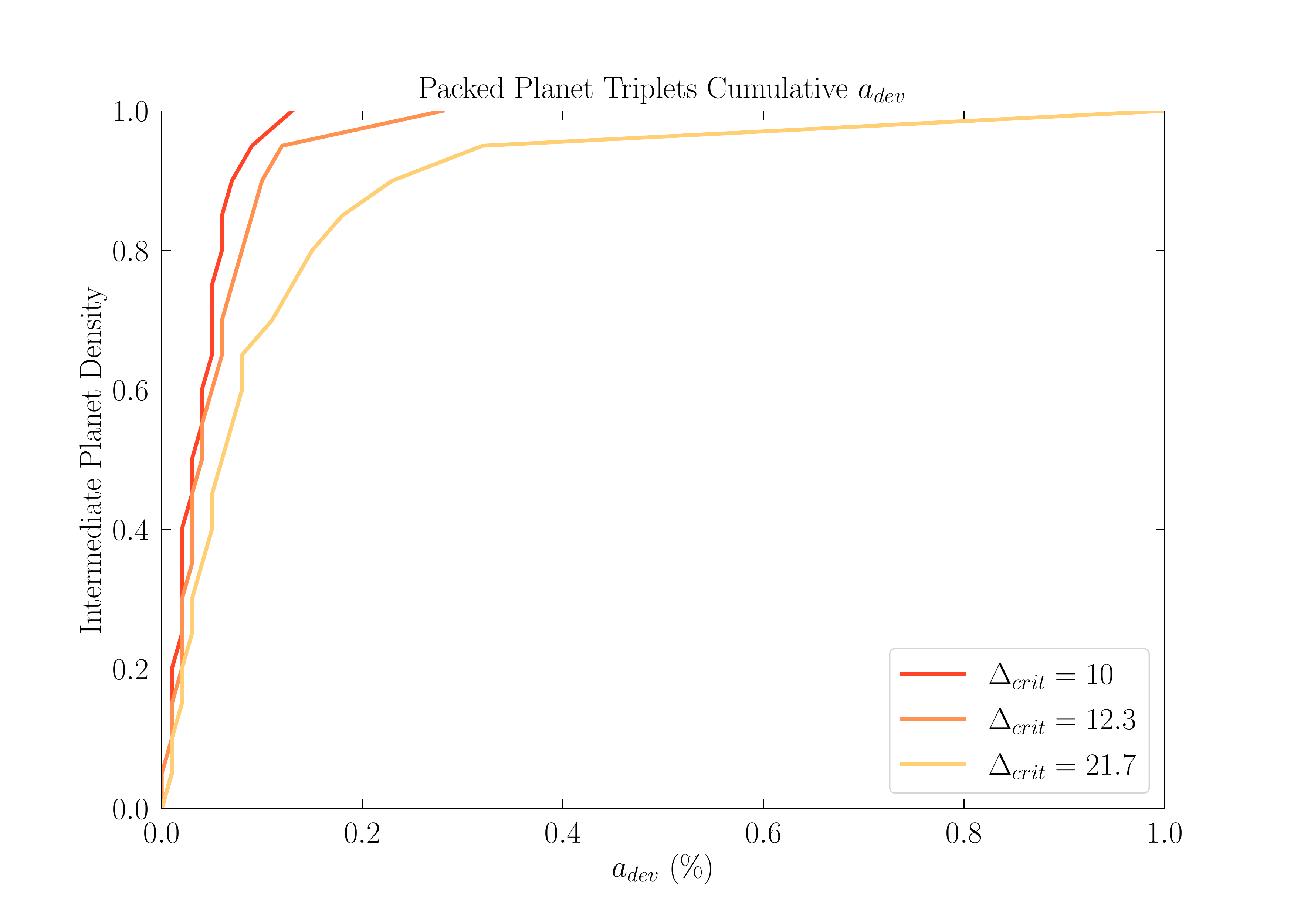}
    \caption{The cumulative $a_{dev}$ distribution. When $\Delta_{crit} = 10$, $12.3$, and $21.7$, $95\%$ of intermediate planet semimajor axes were within $9.44\%$, $12.45\%$ and $24.28\%$ of $a_{ideal}$, respectively.}
    \label{fig:CumulativeAdev}
\end{figure}

All packed triplets in the Kepler multis were identified and the middle planet was removed to simulated unpacked planet pairs. The $M_{eff}$ and $a_{dev}$ were then calculated for all simulated unpacked planet pairs.
The median mass efficiencies in the triplets for $\Delta_{crit} = 10$, $12.3$, and $21.7$ were $42.28\%$, $64.93\%$, and $173.56\%$, respectively (Figure \ref{fig:PackedTripletMassEfficiency}). In some instances, there were intermediate planets with efficiencies that exceeded $100\%$, suggesting either that the Forecaster masses for these planets could be too large or that the systems are unstable. When $\Delta_{crit} = 10$, $95\%$ of intermediate planets were located at a semimajor axis within a $9.44\%$ error of the ideal semimajor axis. When $\Delta_{crit} = 12.3$ and $21.7$, $95\%$ of the intermediate planets had an $a_{dev}$ of less than $12.45\%$ and $24.28\%$, respectively. Thus, the ideal semimajor axis could be a good predictor for the actual semimajor axis of an intermediate planet in a packed triplet (Figure \ref{fig:CumulativeAdev}). 

\begin{figure}
    \centering
    \includegraphics[scale=0.5]{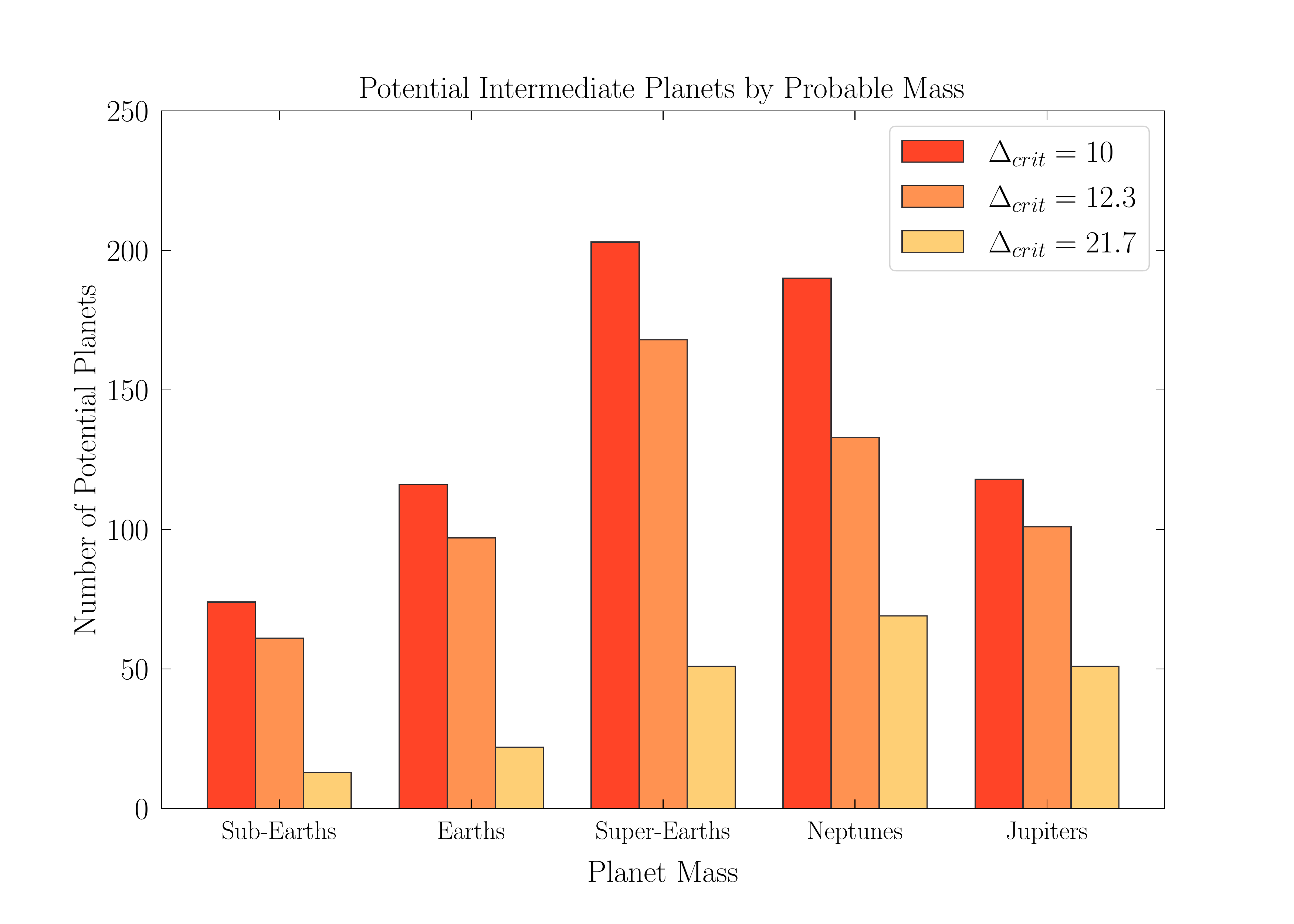}
    \caption{The number of potential intermediate planets by probable masses based on the median $M_{eff}$ for each $\Delta_{crit}$. Sub-Earths have masses less that $0.5$ Earth masses ($M_{\oplus}$), Earths are $0.5-2$ $M_{\oplus}$, Super-Earths are $2-10$ $M_{\oplus}$, Neptunes are $10-50$ $M_{\oplus}$, and Jupiters are $50-5000$ $M_{\oplus}$.}
    \label{fig:PredictionbyMass}
\end{figure}

The results from the packed triplet analysis can be used to predict the parameters of potential undetected planets in unpacked planet pairs. To predict the range of possible masses of a potential planet in an unpacked pair, the minimum, 25th-percentile, median, 75th-percentile, and maximum mass efficiencies for each $\Delta_{crit}$ were multiplied by each planet pair’s mass capacity. If it is assumed that a potential planet hosted by an unpacked planet pair has a mass efficiency equal to the median efficiency, Super-Earths and Neptunes are predicted to be the most common among the potential planets for all three chosen values of $\Delta_{crit}$ (Figure \ref{fig:PredictionbyMass}). It is worth noting that the methods presented here only take into account the possibility of a single missing intermediate planet for each planet planet pair. Thus, it is possible (and likely) that the large ``missing'' masses would be distributed among multiple intermediate planets instead of a single larger planet as predicted here. 

In addition to confirming that $a_{ideal}$  can be used to predict the semimajor axis of potential planets, the semimajor axis deviations calculated for the packed triplets can be used to identify unpacked planet pairs that are most likely to host such additional planets. If given three consecutive planets in a triplet containing at least one unpacked planet pair, the middle planet can be removed to calculate $a_{dev}$ using the same method as the packed triplet analysis. If the middle planet’s $a_{dev}$ is sufficiently large, it could suggest that an additional planet might exist in the unpacked planet pair of which the middle planet is a part that is pushing it away from its expected $a_{ideal}$. Assuming $\Delta_{crit}=12.3$ and defining “sufficiently large” as any $a_{dev}$ greater than $12.45\%$ (beyond the $a_{dev}$ of $95\%$ of the middle planets in the packed triplet analysis), there were $96$ unpacked Kepler triplets with a heightened probability of containing an additional planet.

\section{Discussion}\label{sec:discussion}
The dynamical structures of all Kepler multi-candidate systems were assessed to identify unpacked planet pairs capable of hosting intermediate planets and determine the upper limits for these missing intermediate planets. Possible masses and semimajor axes then were constrained for these unidentified planets. Overall, Kepler multi-candidate systems could host as many as $701$ ($\Delta_{crit}=10$), $560$ ($\Delta_{crit}=12.3$), or 206 ($\Delta_{crit} = 21.7$) additional planets. Earths and Sub-Earths combined could make up $27.10\%$, $28.21\%$, or $16.99\%$ of these planets according to $\Delta_{crit} = 10$, $12.3$, and $21.7$, respectively. Super-Earths are predicted to be the most common among the potential planets.

There are two populations of potential missing planets that should be targeted in follow-up investigations. The first population lies within systems with three or more known planets. When three or more planets exist in a system, intermediate planet(s) can be ``removed'' one at a time and their ideal semimajor axis can be predicted using the solution to the system of Equations \ref{axmin} and \ref{axmax} in a manner similar to the packed triplet analysis conducted in Section \ref{sec:predicting}. The removed planet's predicted ideal semimajor axis can then be compared to its actual semimajor axis and its $a_{dev}$ can be calculated. In Kepler packed triplets, $95\%$ of intermediate planets had an $a_{dev}$ less than $9.44\%$, $12.45\%$, and $24.28\%$ when $\Delta_{crit} = 10$, $12.3$, and $21.7$, respectively, suggesting that planets in packed systems prefer to orbit in ``gravitational valleys'' where forces are balanced from neighboring planets. Thus, if a removed planet's $a_{dev}$ exceeds these thresholds and the planet is part of an unpacked planet pair, it could suggest that a missing planet exists in the gap within the planet pair. The presence of an additional planet within the planet pair would shift the ideal semimajor axis of the removed planet, thus reconciling the ideal semimajor axis with the planer's observed semimajor axis. In a follow-up analysis, this phenomenon was identified in $96$ planet pairs, indicating a higher likelihood that an unidentified planet is present in these pairs. 

The second population that should be targeted is planets close to the stars predicted to have a radius less than 1.2 earth radii as these smaller planets are easier to miss and can fit in smaller regions of gravitational stability between planet pairs. Planet pairs closest to their star (1-2 Pair) were most likely to be unpacked. When $\Delta_{crit}=12.3$, $50\%$ to $70\%$ of 1-2 Pairs were unpacked. When $\Delta_{crit} = 10$, $60\%$ to $80\%$ of 1-2 Pairs were unpacked in all systems except six-candidate systems. Yet Kepler is more likely to detect transits of planets closer to a host star because the planets would have completed more orbits during the observation period and have had a wider range of detectable inclinations \citep{Thompson2018}. If unidentified planets exist within unpacked 1-2 Pairs, they must have orbits inclined out of the plane of view or a small radius producing a shallow light curve with a low signal-to-noise ratio relative to the host star’s variations in brightness. As concluded in \cite{Howell2016}, Kepler had difficulty detecting planets less than 1.2 Earth radii. Thus, if planets do exist in these gaps, it is very possible that they are smaller planets beyond the detection capabilities of Kepler.

Finding the potential planets suggested in this investigation would support the PPS hypothesis by demonstrating extrasolar systems tend toward a dynamically packed architecture. Not finding these planets could suggest the need to modify PPS, look to alternate theories to describe extrasolar system architecture, or develop different theories. The masses and semimajor axes found in this investigation could help researchers anticipate potential planets’ transit and radial velocity signals, thus enabling researchers to conduct more targeted searches with greater sensitivity to detections with low signal-to-noise ratios.

It is possible that the unpacked planet pairs could be explained by alternate hypotheses regarding system architectures. For example, systems may start packed and then become unstable over time as hypothesized by \cite{PuandWu2015}. It is also plausible that PPS only describes a subpopulation of the Kepler multiplanet systems and that other subpopulations exist that are described by different architectures due to different formation processes or histories of dynamical evolution. These possibilities could be explored through long-term integrations of Kepler systems to study how the systems evolve or applying analyses based in information theory as done by \cite{Kipping2018}.

Uncertainties in the data serve as the biggest limitation in this investigation. Uncertainties in stellar parameters are a significant source of error for planet parameters as the planet parameters are calculated from stellar parameters \citep{Burke2015}. The uncertainties in stellar mass and radius are most concerning as the stellar mass is a direct input for dynamical spacing and stellar radius is used to determine planetary radius, which is then converted to planetary mass, another direct input. Uncertainties in stellar properties are about $27\%$ for stellar radius and $17\%$ for stellar mass \citep{Thompson2018}. Furthermore, unlike the error that comes from the uniform Kepler pipeline, the uncertainties in stellar parameters are non-uniform across the sample as the data is compiled from many different astronomical surveys and instruments. Recently, \cite{Weiss2018} made efforts to better characterize the stellar parameters of a large population of stars observed by the original Kepler mission to improve the accuracy of planetary parameters. As stellar parameters continue to be updated, the upper limits for potential missing planets presented here can be updated as well.

Error can also be caused by background light from stars during transit detections as the light can wash out transit signals and make them appear shallower, causing the under-estimation of planet radii \citep{Thompson2018}. This error is non-uniform across the sample and can only be eliminated by careful follow-up characterizations of planet parameters with other instruments. If radii are underestimated, then the mass of the planets would be underestimated as well. If some planets are indeed more massive, then a larger portion of the planet pairs will likely be packed and fewer planet pairs will be capable of hosting additional planets. Calculating the mass of planets from the radius using an empirical mass-radius relation can also provide error as the relation is not absolute as planets can have varying densities \citep{ChenandKipping2017}. Results from the TESS mission, which had a primary goal to find at least 50 planets smaller than Neptune that orbit stars bright enough to enable follow-up mass measurements, will help constrain mass-radius relations.
Another limitation to this investigation is that the equations used here assume that the multi-candidate systems are fairly coplanar and contain planets with circular obits. This concern is partially addressed in this investigation by assuming $\Delta_{crit}=12.3$, a slightly larger dynamical spacing resulting from simulations with slight variations in eccentricity and inclination \cite{PuandWu2015}. However, this limitation could be further addressed by employing equations similar to those used by \cite{Gladman1993}, which directly take into account orbital eccentricity.

\section{Conclusion}\label{sec:Conclusion}
Many Kepler multi-candidate systems have space for additional intermediate planets. The analytical examination shows that Kepler multi-candidate systems may have as many as $701$ ($\Delta_{crit}=10$), $560$ ($\Delta_{crit}=12.3$), or $206$ ($\Delta_{crit}=21.7$) additional planets. The MC analysis shows that $209$ ($\Delta_{crit}=10$), $164$ ($\Delta_{crit}=12.3$), and $118$ ($\Delta_{crit}=21.7$) planet pairs have a probability $\geq0.90$ of being unpacked.
Many potential planets could be in Earth’s mass neighborhood. The MC analysis suggests that $M_{xmax}$ of as many as $370$, $487$, and $523$ planet pairs’ mean is $\leq 10 M_{\oplus}$ (Super-Earth or smaller) when $\Delta_{crit}=10$, $12.3$, and $21.7$, respectively. According to median $M_{eff}$, $27.10\%$, $28.21\%$, or $16.99\%$ of these planets could be Earths and Sub-Earths when $\Delta_{crit}=10$, $12.3$, and $21.7$, respectively. However, the large tail of the MC mean $M_{xmax}$  distribution indicates that there is space in some pairs for larger planets too.
Planets in packed systems tend to orbit at or near their $a_{ideal}$. In Kepler packed triplets, $95\%$ of planets had an $a_{dev} \leq 9.44\%$, $12.45\%$, and $24.28\%$ when $\Delta_{crit}=10$, $12.3$, and $21.7$, respectively. This suggests that planets tend towards orbits where gravitational forces are most balanced (“gravitational valleys”).
In conclusion, we present a method for analytically analyzing the dynamical packing of multiplanet systems that can be applied easily and rapidly to large populations of systems. The wider application to currently known multiplanet systems as well as incoming systems from missions such as TESS could provide insight on potentially undetected low-mass planets as well as the systems’ inherent architectures and formation histories.

\section{Acknowledgements}\label{sec:Acknowledgements}
We would like to thank Dimitar Sasselov, Zoe Todd, and Veselin Kostov for their thorough comments and suggestions during the preparation of this paper and Serena Wurmser for her figure formatting recommendations. We would also like to thank Mark Otto and Tom Barclay for their many helpful insights and discussions over the course of this research. We would like to acknowledge the following individuals who have provided feedback and support at various stages during the research process: Rory Barnes, Jonathan McDowell, Saul Rappaport, and Alan Gould. Finally, we would like to thank Shawn Lowe who provided critical guidance and instruction during this project and without whom this research would not have been possible.



\end{document}